\documentclass[aps,prb,twocolumn,superscriptaddress]{revtex4-2}

\usepackage{bm}% bold math
\usepackage{lipsum}% http://ctan.org/pkg/lipsum
\usepackage{graphicx,amssymb}
\usepackage{dcolumn}% Align table columns on decimal point
\usepackage{bm}% bold math
\usepackage{xcolor}
\usepackage{amsmath}
\usepackage{hyperref}

\begin{document}

\title{Observation of Subharmonic Charge-Density-Wave Correlations in La-Based Cuprates}

\author{J.-S. Lee}\email[]{jslee@slac.stanford.edu}
\affiliation{Stanford Synchrotron Radiation Lightsource, SLAC National Accelerator Laboratory, Menlo Park, California 94025, USA}
\author{S. A. Kivelson}
\affiliation{Departments of Physics, Stanford University, Stanford, CA 94305, USA}
\author{H. Lee}
\affiliation{Stanford Synchrotron Radiation Lightsource, SLAC National Accelerator Laboratory, Menlo Park, California 94025, USA}
\author{T. Wang} 
\affiliation{Institute for Materials Research, Tohoku University, Katahira 2-1-1, Sendai, 980-8577, Japan}
\author{Y. Ikeda} 
\affiliation{Institute for Materials Research, Tohoku University, Katahira 2-1-1, Sendai, 980-8577, Japan}
\author{T. Taniguchi} 
\affiliation{Institute for Materials Research, Tohoku University, Katahira 2-1-1, Sendai, 980-8577, Japan}
\author{C.-T. Kuo}
\affiliation{Stanford Synchrotron Radiation Lightsource, SLAC National Accelerator Laboratory, Menlo Park, California 94025, USA}
\author{M. Fujita}
\affiliation{Institute for Materials Research, Tohoku University, Katahira 2-1-1, Sendai, 980-8577, Japan}
\author{C.-C. Kao}
\affiliation{Stanford Synchrotron Radiation Lightsource, SLAC National Accelerator Laboratory, Menlo Park, California 94025, USA}
\affiliation{SLAC National Accelerator Laboratory, Menlo Park, California 94025, USA}

\begin{abstract}
Pair-density-wave (PDW) correlations have been proposed as an important ingredient in the complex phase diagram of high-$T_{\rm c}$ cuprates, yet bulk-sensitive experimental signatures remain scarce. Here we report resonant x-ray scattering measurements revealing a subharmonic charge-density-wave (CDW) scattering response in Sr-doped $1/8$-LBCO. The subharmonic response appears at approximately half the primary CDW ordering wave vector and emerges within a physically relevant temperature regime associated with the development of in-plane superconducting correlations. Comparable subharmonic behavior is also observed in a chemically distinct La-based cuprate, LSCO, within the same stripe-ordered, layer-decoupled regime. Together, these observations identify a bulk-sensitive scattering signature that is consistent with PDW correlations in La-based superconducting cuprates.
\end{abstract}

\maketitle

\section{Introduction}

Numerous studies have reported evidence suggestive of a novel form of superconducting order, the pair-density-wave (PDW), appearing over restricted ranges of doping and temperature in several high-$T_{\rm c}$ cuprates~\cite{Tajima1,Yuli1,Li1,Yang1,Yuli2,Schafgans1,Stegen1,Hamidian1,Edkins1,Du1,Huang1,Lozano1,Lozano2,Lee1,Dai1,Lee2,Agterberg1,Zelli1,Yu1,Norman1,Himeda1,Berg1,Berg2,Berg3,Radzihovsky1,Agterberg2,Fradkin1,Mross1,Cai1,Wang1,Ding1,Shi1,Zhong1,Agterberg3}. A key early indication came from the 2007 observation by Li \textit{et al.} in La$_{1.875}$Ba$_{0.125}$CuO$_4$ (LBCO) of pronounced layer decoupling—superconducting behavior confined to individual CuO$_2$ planes in the apparent absence of interlayer Josephson coupling~\cite{Li1}. Subsequent transport studies reported similar phenomenology in other La-based cuprates, including La$_{2-x-y}X_y$Sr$_x$CuO$_4$ with $X=$ Nd, Eu, and Ca~\cite{Ding1,Shi1,Zhong1}, Fe- and Zn-doped La$_{2-x}$Sr$_x$CuO$_4$~\cite{Huang1,Lozano1}, and underdoped La$_{2-x}$Sr$_x$CuO$_4$ in applied magnetic fields~\cite{Schafgans1}, lending further support to the relevance of PDW correlations.

A major advance was provided by scanning tunneling microscopy measurements on Bi$_2$Sr$_2$CaCu$_2$O$_{8+\delta}$ (Bi2212), where Edkins \textit{et al.} observed a subharmonic modulation of the charge-density-wave (CDW) order in the vicinity of magnetic-field-induced vortices~\cite{Edkins1}. The subharmonic ordering vector, $\vec{K}/2$, is half that of the primary CDW order $\vec{K}$ and is naturally expected in the presence of coexisting PDW correlations with wave vector $\vec{1Q}=\vec{K}/2$ and uniform superconductivity~\cite{Edkins1,Dai1,Agterberg1,Wang1,footnote1}. Similar considerations apply more generally to regimes in which PDW and/or uniform superconducting correlations are fluctuating but retain sufficiently long correlation lengths.

Despite these advances, the broader implications of PDW-related signatures remain unsettled. In Bi2212, the CDW correlations are short-ranged, with correlation lengths only marginally exceeding the CDW period~\cite{Neto1,Lu1,Chaix1}, information on spin-density-wave order at zero field is limited~\cite{Mounce1}, and no macroscopic manifestations such as layer decoupling have been observed. More generally, no bulk-sensitive diffraction measurements have reported subharmonic signatures in La-based cuprates, Bi2212, YBa$_2$Cu$_3$O$_{6+x}$, or other cuprate families, even in regimes where such features might be anticipated if PDW correlations were ubiquitous~\cite{Chaix1,Vinograd,Blackburn1,Peng1,Schafgans}.

In this study, we investigate subharmonic charge-density-wave responses in two La-based cuprates near 1/8 doping, Sr-doped La$_{1.875}$Ba$_{0.125}$CuO$_4$ and Fe-doped La$_{1.87}$Sr$_{0.13}$CuO$_4$, using resonant soft x-ray scattering. Transport measurements are used to establish the temperature regime associated with superconducting correlations. In both materials, the subharmonic response emerges as an additional RSXS intensity near the putative subharmonic wave vector ($\vec{1Q}$) upon cooling into the regime where the in-plane resistivity $\rho_{\rm ab}$ exhibits a superconducting-like drop, while the interplane resistivity $\rho_{\rm c}$ remains finite, indicating the absence of three-dimensional superconducting long-range order. The phrase ``three-dimensional superconducting long-range order'' is used here operationally to denote bulk interlayer phase coherence, rather than to imply that superconducting pairing itself requires three-dimensional coupling. These observations provide a bulk-sensitive scattering signature consistent with PDW correlations in La-based cuprates.

\section{Experiments}

\subsection{Materials and Methods}

Two high-quality single crystals, Sr-doped LBCO (La$_{1.875}$Ba$_{0.065}$Sr$_{0.06}$CuO$_4$) and Fe-doped LSCO (La$_{1.87}$Sr$_{0.13}$Cu$_{0.99}$Fe$_{0.01}$O$_4$), are employed for this study. Both crystals were grown by the traveling-solvent floating-zone method. After growth, the crystals were annealed in one bar of O$_2$ gas to minimize oxygen deficiency. The typical growth rate was 1.0~mm/hr, yielding crystal rods with lengths of approximately 50--60~mm for each composition. Sr substitution in LBCO and Fe substitution in LSCO perturb stripe correlations through different microscopic routes. Fe substitutes directly in the CuO$_2$ planes and has been reported to enhance or pin spin stripe correlations through its magnetic character \cite{Suzuki,FujitaLSCFO}, whereas Sr substitution outside the CuO$_2$ planes more directly modifies the lattice environment and structural tendencies that influence stripe order and interlayer coupling \cite{Fujita1,Fujita2}. The present study focuses on the common emergence of a subharmonic scattering response in the layer-decoupled regime across two La-based cuprates with different substitution routes, rather than separating the microscopic effects of the two substitutions.

Resonant soft x-ray scattering experiments were carried out at beamline 13-3 of the Stanford Synchrotron Radiation Lightsource (SSRL). Initial measurements were performed using an in-vacuum four-circle diffractometer equipped with an open-cycle helium cryostat, providing a base temperature of approximately 25~K. Subsequent measurements in 2024 employed an upgraded RSXS instrument with a closed-cycle helium cryostat, enabling measurements down to a base temperature of approximately 12~K~\cite{Kuo1}. Because these measurements were performed using different instrumental configurations and independent sample cleaves, we analyze temperature-dependent changes within each data set and do not use absolute intensities from different configurations for quantitative comparison.

For RSXS experiments, samples were prepared with typical dimensions of 1.5~mm $\times$ 1.5~mm $\times$ 2.5~mm along the $a$, $b$, and $c$ axes, respectively. To obtain a fresh $c$-axis-normal surface [aiming to align the scattering plane along the ac-plane, each sample was cleaved \emph{ex situ} immediately prior to transfer into the ultra-high-vacuum chamber (base pressure $\sim 8 \times 10^{-10}$~Torr). Additional details are provided in the Supplemental Material~\cite{Supple}. Separate crystals with lengths of approximately 25~mm were used for elastic neutron scattering measurements to characterize the spin-density-wave (SDW) ordering temperature in Sr-doped LBCO.

\subsection{Intertwined density waves}

\begin{figure}[t]
\begin{center}
\includegraphics[width=0.5\textwidth]{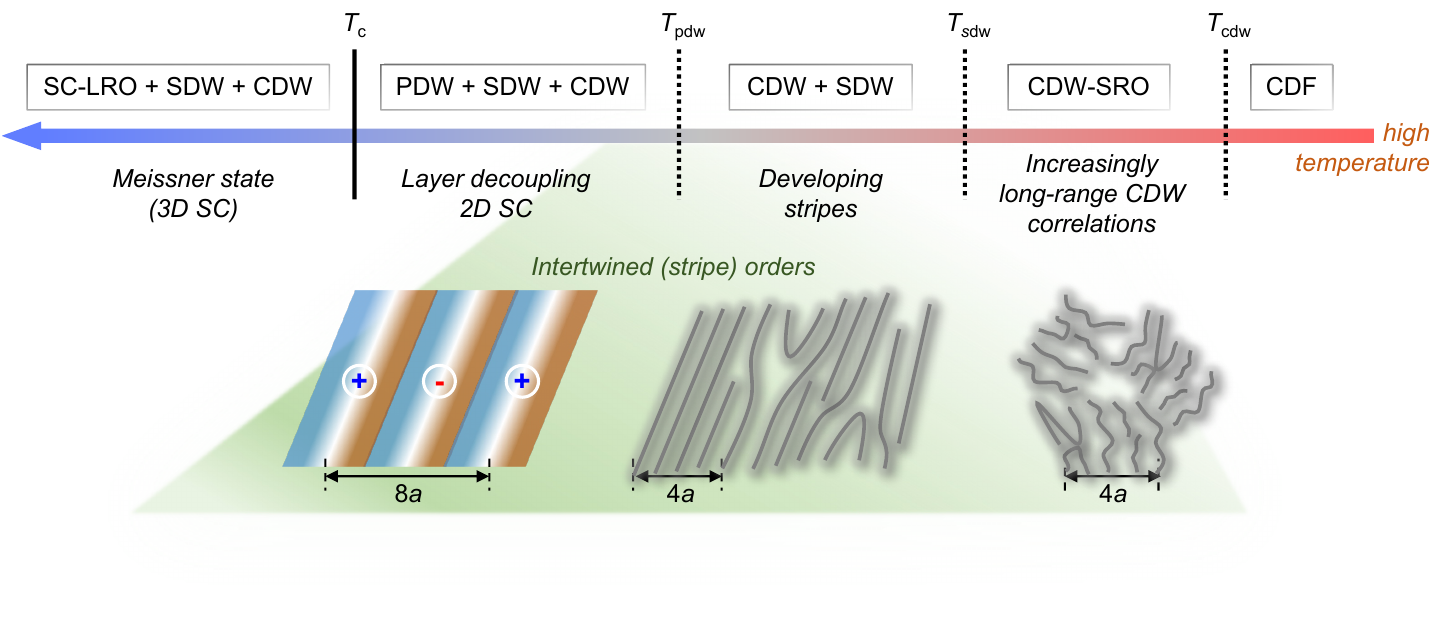}
\caption{Schematic illustration of the hierarchy of intertwined density-wave and superconducting correlations in La-based cuprates. Upon cooling from high temperatures, fluctuating charge-density-wave order (CDF) \cite{Arpaia1} develops first and subsequently evolves into short-range CDW order (CDW-SRO), eventually giving rise to a stripe-ordered state characterized by coexisting CDW and SDW correlations \cite{Wen1,Wen2}. Within this stripe-ordered background, additional electronic correlations, such as PDW correlations, may emerge at intermediate temperatures \cite{Li1,Huang1}, preceding the establishment of three-dimensional superconducting long-range order. The lower panel shows a real-space cartoon illustrating the corresponding intertwined orders, highlighting that a PDW modulation can naturally allow a doubling of the charge modulation period from 4$a$ to 8$a$, corresponding to a subharmonic component relative to the primary CDW.} 
\label{Fig1}
\end{center}
\end{figure}

The La-based cuprates studied here host multiple intertwined density-wave (DW) orders, including CDW and spin-density-wave (SDW) orders, that are already well established over a broad temperature range~\cite{Li1,Huang1,Fradkin1}. Figure~\ref{Fig1} schematically illustrates a hierarchy of such intertwined orders. Upon cooling from high temperatures, short-range CDW correlations first develop and evolve into a stripe-ordered state characterized by coexisting CDW and SDW correlations with long correlation lengths. These stripe correlations provide the broader physical background for the present observations. Within this stripe-ordered background, additional electronic correlations may emerge at lower temperatures, preceding the establishment of three-dimensional superconducting long-range order. This hierarchy provides a natural framework for investigating subharmonic scattering signatures associated with PDW correlations in the intermediate temperature regime.

%%%%%%%%%%%%%%%%Fig1%%%%%%%%%%%%%%%%%%%%%%%%%%%%%%
%%%%%%%%%%%%%%%%%%%%%%%%%%%%%%%%%%%%%%%%%%%%%%

In La-214 stripe systems, the CDW and SDW ordering wave vectors are known to be harmonically related, with \(q_{\rm CDW}\simeq 2q_{\rm SDW}\)~\cite{yamadarule1,yamadarule2,yamadarule3}. Therefore, a response near \(q_{\rm CDW}/2\) occurs at a wave vector close to the SDW ordering wave vector. This relation is an important part of the stripe phenomenology and must be distinguished from the present interpretation. The signal discussed here is observed in the Cu \(L_3\)-edge resonant scattering channel as a weak charge-related subharmonic response relative to the primary CDW wave vector. We therefore do not interpret the feature simply as the known SDW order, but rather as a resonant subharmonic scattering response appearing within the stripe-ordered, layer-decoupled regime. In this context, we explore subharmonic scattering signatures associated with PDW correlations within the stripe-ordered regime using RSXS. In such a setting, interference between PDW and uniform superconducting correlations is expected to allow a subharmonic scattering response at half the primary CDW wave vector, commonly referred to as a 1Q feature.

\subsection{Subharmonic feature on Sr-doped LBCO}

\begin{figure}[t!]
\begin{center}
\includegraphics[width=0.48\textwidth]{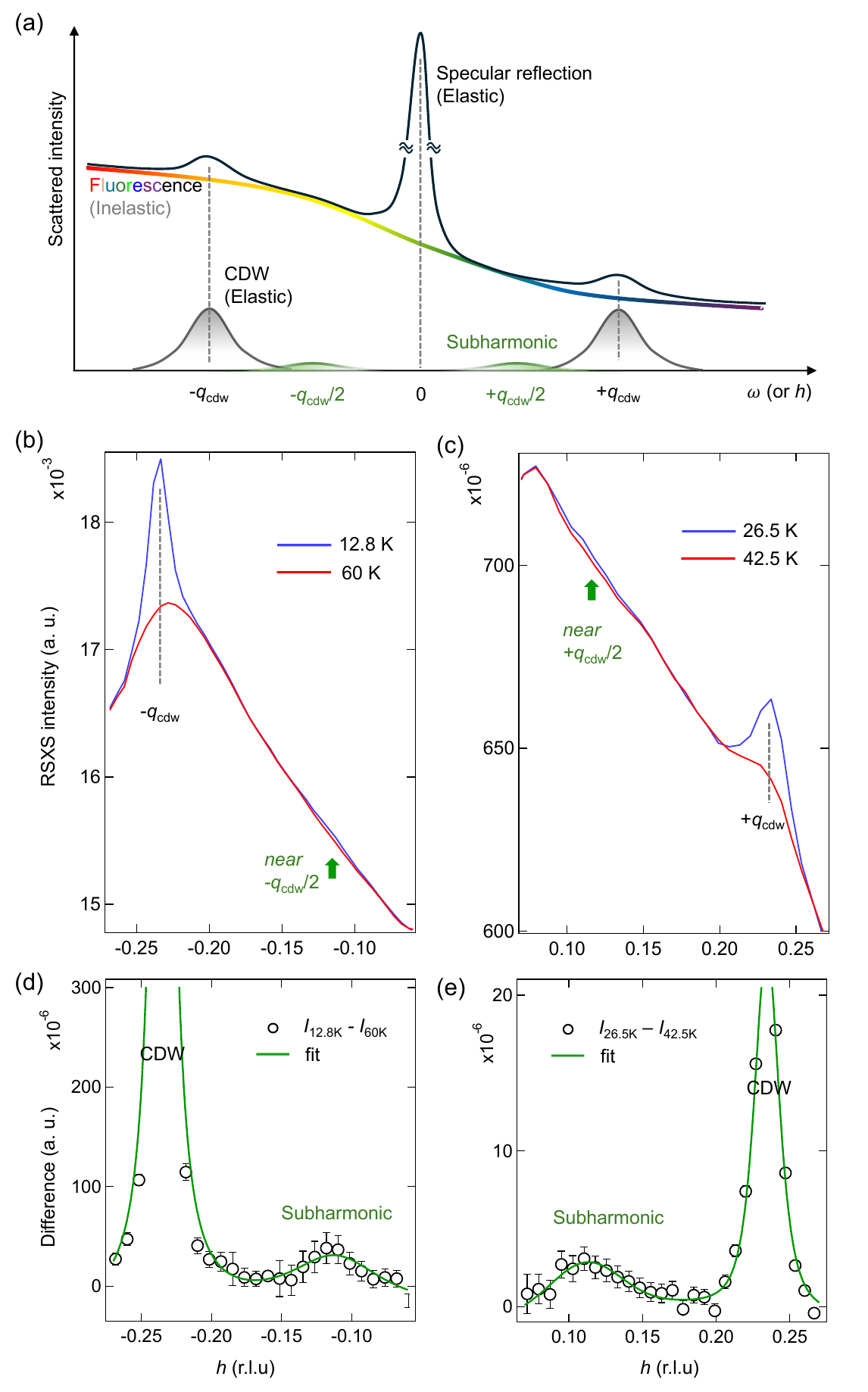}
\caption{Subharmonic scattering response in Sr-doped LBCO revealed by RSXS.
(a) Schematic illustration of the momentum regions probed in the RSXS measurements, indicating the elastic CDW peak at $\pm q_{\rm cdw}$, the subharmonic region near $\pm q_{\rm cdw}/2$, fluorescence background, and the specular reflection. (b,c) Representative momentum-dependent RSXS profiles measured along the CDW wave-vector direction at two temperatures in the stripe-ordered regime. While the primary CDW peak is clearly visible, a subtle temperature-dependent difference appears near the subharmonic positions.  Note the different vertical scales in panels (b) and (c). (d,e) Difference between the low- and high-temperature RSXS intensities shown in (b) and (c), respectively. In addition to the dominant contribution at the primary CDW wave vector, a residual temperature-dependent signal is present near the subharmonic positions. The solid fit lines are guides to the eye used to indicate the approximate momentum location of the temperature-dependent contribution near $\pm q_{\rm cdw}/2$. The statistical uncertainty is approximately 1$\times$10$^{-5}$ a.u., corresponding to a signal that exceeds the uncertainty by several times.} \label{Fig2}
\end{center}
\end{figure}

To investigate this possibility, we performed RSXS experiments on Sr-doped La$_{1.875}$Ba$_{0.125}$CuO$_4$ (LBCO) \cite{Supple}, which has played a central role in the development of the PDW concept. LBCO is an archetypal system in which stripe order, layer-decoupled superconductivity, and anomalous transport signatures have long suggested the presence of PDW correlations \cite{Li1}. As shown in Fig.~\ref{Fig2}(a), Cu-resonant scattering measurements in this regime are complicated by substantial background contributions from fluorescence and specular reflection at the sample surface. In addition, a pronounced static CDW signal can further hinder the experimental resolution of subharmonic scattering components, motivating the choice of lightly Sr-doped LBCO in the present study. While pure 1/8-doped LBCO is expected to host strong stripe and PDW correlations, the associated CDW scattering is so dominant that any subharmonic contribution may remain experimentally buried in RSXS. Partial Sr substitution provides a controlled reduction of the CDW background, thereby enhancing the visibility of subharmonic scattering components without qualitatively altering the underlying stripe-ordered regime.

%%%%%%%%%%%%%%%%Fig2%%%%%%%%%%%%%%%%%%%%%%%%%%%%%%
%%%%%%%%%%%%%%%%%%%%%%%%%%%%%%%%%%%%%%%%%%%%%%

Figures~\ref{Fig2}(b) and (c) show representative momentum-dependent RSXS profiles measured at both $\pm q_{\rm cdw}$. Pronounced CDW peaks are clearly observed near the characteristic $4a$ periodicity, corresponding to $h \approx \pm 0.235 \pm 0.005$~r.l.u. In addition to the primary CDW signal, the raw RSXS profiles reveal an intensity near the tails of the CDW peaks on both sides of the CDW peak positions ($h > -0.15$ and $h < +0.15$~r.l.u.), as indicated by the arrows near $\pm q_{\rm cdw}/2$. This suggests that an additional subharmonic scattering component is already discernible in the raw data within the statistical sensitivity of the measurement.

In order to isolate temperature-dependent scattering contributions beyond the dominant CDW background, and to suppress temperature-independent backgrounds such as fluorescence and specular scattering, we compare data acquired at low temperature with those measured above the SDW ordering temperature $T_{\rm sdw} \approx 35.4$~K~\cite{Supple,Fujita1,Fujita2}. The resulting difference profiles, shown in Figs.~\ref{Fig2}(d) and (e), highlight residual features beyond the dominant CDW contribution. These profiles reveal an excess scattering intensity near $h \approx \pm 0.12 \pm 0.007$~r.l.u. below $T_{\rm sdw}$, consistent with a putative subharmonic response. We note that a weak structure near $h$ $\sim$ +0.08~r.l.u. is visible in the raw profile of Fig. 2(c). This structure is not used in the analysis or interpretation of the subharmonic response, which is instead identified from the temperature-dependent contribution near $+q_{\rm cdw}/2$ in the difference profile.

%%%%%%%%%%%%%%%%Fig3%%%%%%%%%%%%%%%%%%%%%%%%%%%%%%
\begin{figure}[b!]
\begin{center}
\includegraphics[width=0.5\textwidth]{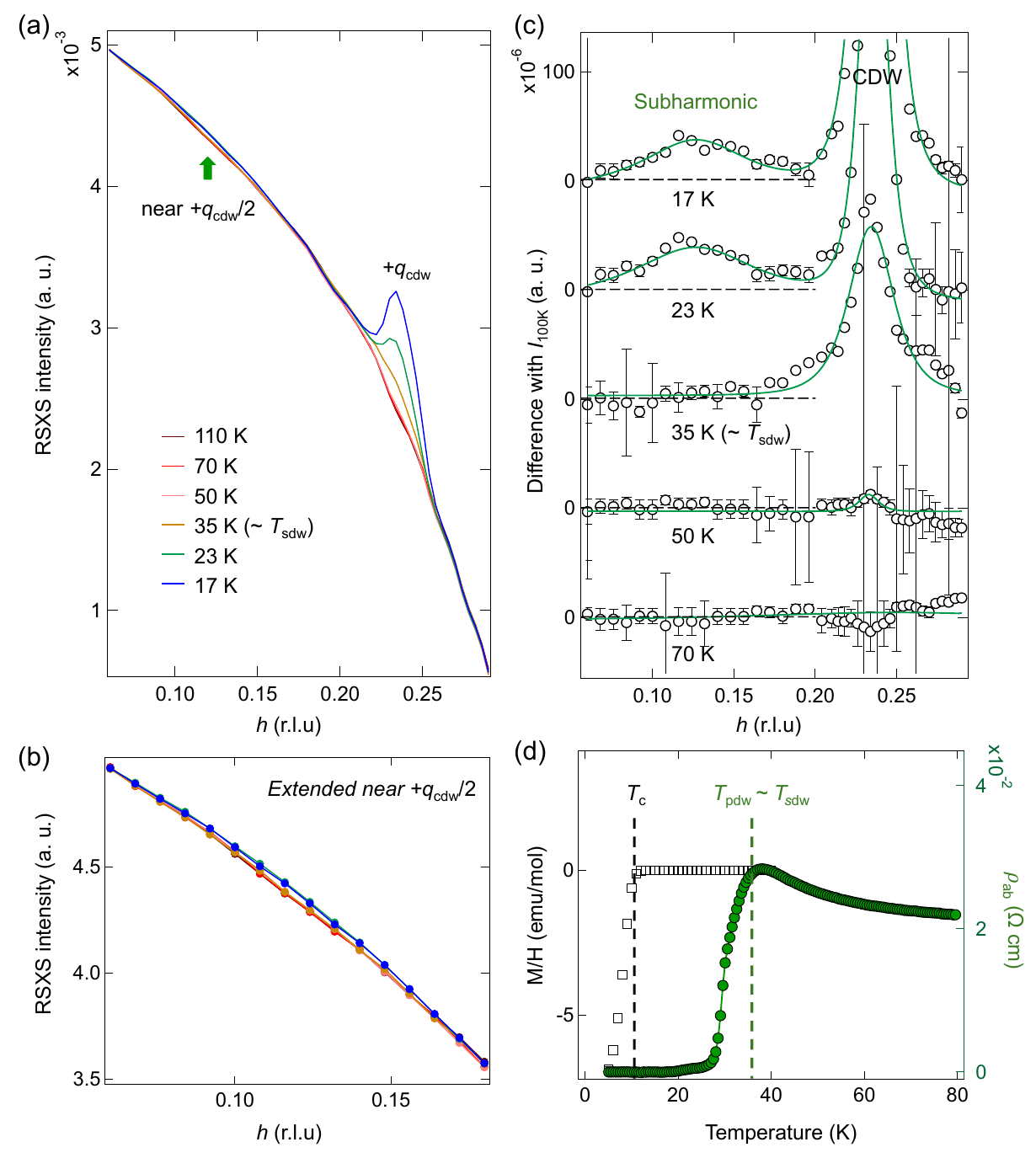}
\caption{Temperature dependence of the subharmonic scattering response and in-plane transport in Sr-doped LBCO. (a) Raw momentum-dependent RSXS profiles measured along the CDW wave-vector direction at selected temperatures. The crystal was re-cleaved before the temperature-dependent measurements, and the data shown here represent an independent data set from Fig.~\ref{Fig2}. (b) Expanded view of the RSXS intensity near the subharmonic region around $+q_{\rm cdw}/2$, illustrating the temperature evolution of scattering features superimposed on the dominant CDW background. (c) Temperature-difference spectra obtained by subtracting the RSXS profile measured at 110~K from those measured at lower temperatures, highlighting temperature-dependent scattering components near the subharmonic wave vector. Solid lines represent simple fits used as guides to the eye. (d) Superconducting properties as a function of temperature, showing the bulk magnetic susceptibility (left \textit{y}-axis, open squares) and the zero-field in-plane resistivity $\rho_{\rm ab}$ (right \textit{y}-axis, green filled circles)}.
\label{Fig3}
\end{center}
\end{figure}
%%%%%%%%%%%%%%%%%%%%%%%%%%%%%%%%%%%%%%%%%%%%%%

Given that the subharmonic response appears as a subtle modification of the dominant CDW scattering profile and is experimentally buried beneath the strong CDW contribution and background scattering, its characterization is not fully captured by peak-based metrics alone. In this regard, we examine its temperature dependence to clarify the temperature range over which this response becomes discernible. 

Figure~\ref{Fig3}(a) summarizes the temperature evolution of the RSXS response in the vicinity of both the primary CDW wave vector and the subharmonic wave vector. Upon cooling, an excess RSXS intensity near the subharmonic wave vector, relative to higher-temperature data, becomes discernible below 35 K [Fig.~\ref{Fig3}(b)], while the primary CDW signal gradually weakens with increasing temperature but remains the dominant contribution. By comparing these data with measurements taken at a higher reference temperature (110~K) through a temperature-difference analysis [Fig.~\ref{Fig3}(c)], we identify a residual subharmonic response that becomes evident below the SDW ordering temperature $T_{\rm sdw}$. Note that weak residual side structures near the CDW peak at 35 K become suppressed upon cooling even as the CDW peak grows, indicating that they do not scale with the CDW intensity. Such minor residual structures are consistent with subtle temperature-dependent background or line-shape variations, which may include changes in the fluorescence contribution. Importantly, this temperature window coincides with the regime in which the in-plane resistivity $\rho_{\rm ab}$ exhibits a pronounced superconducting-like reduction [Fig.~\ref{Fig3}(d)], while the interplane resistivity $\rho_{\rm c}$ remains finite, indicating the absence of three-dimensional superconducting long-range order down to approximately 10~K ($T_{\rm c}$). This correspondence indicates that the subharmonic response becomes discernible specifically within the regime of in-plane superconducting correlations, rather than tracking the overall strength of the CDW order. This temperature correspondence should be distinguished from a simple structural-transition effect. Previous studies of Sr-doped LBCO have shown that the LTO/LTT/LTLO structural evolution strongly affects stripe order, particularly the primary CDW component, which is stabilized in the LTT/LTLO phases~\cite{Fujita2}. At the same time, SDW correlations are comparatively more robust against changes in the structural phase, with substantial SDW intensity persisting even toward the LTO regime. Thus, while the structural transition provides an important lattice environment for stripe formation, it does not by itself uniquely determine the subharmonic response observed here. In the present data, the response near $q_{\rm cdw}/2$ becomes discernible in the regime where SDW correlations and layer-decoupled superconducting transport signatures develop, rather than simply tracking the overall primary CDW intensity.

\subsection{Subharmonic feature on Fe-doped LSCO}

Having established the phenomenology identified in Sr-doped LBCO under conditions where stripe-ordered charge correlations are well defined, we now ask whether a detectable subharmonic imprint can also be identified in a La-based compound, such as La$_{2-x}$Sr$_x$CuO$_4$, where such correlations are less robust, in the regime where signatures of two-dimensional superconducting correlations have been reported. We extend the study to Fe-doped La$_{1.87}$Sr$_{0.13}$CuO$_4$ (LSCO)~\cite{Supple}, which provides a natural counterpoint to LBCO. In particular, previous transport and scattering studies on this cuprate reported signatures consistent with interlayer decoupling in this composition range~\cite{Huang1,FujitaLSCFO}.

\begin{figure}[t!]
\begin{center}
\includegraphics[width=0.49\textwidth]{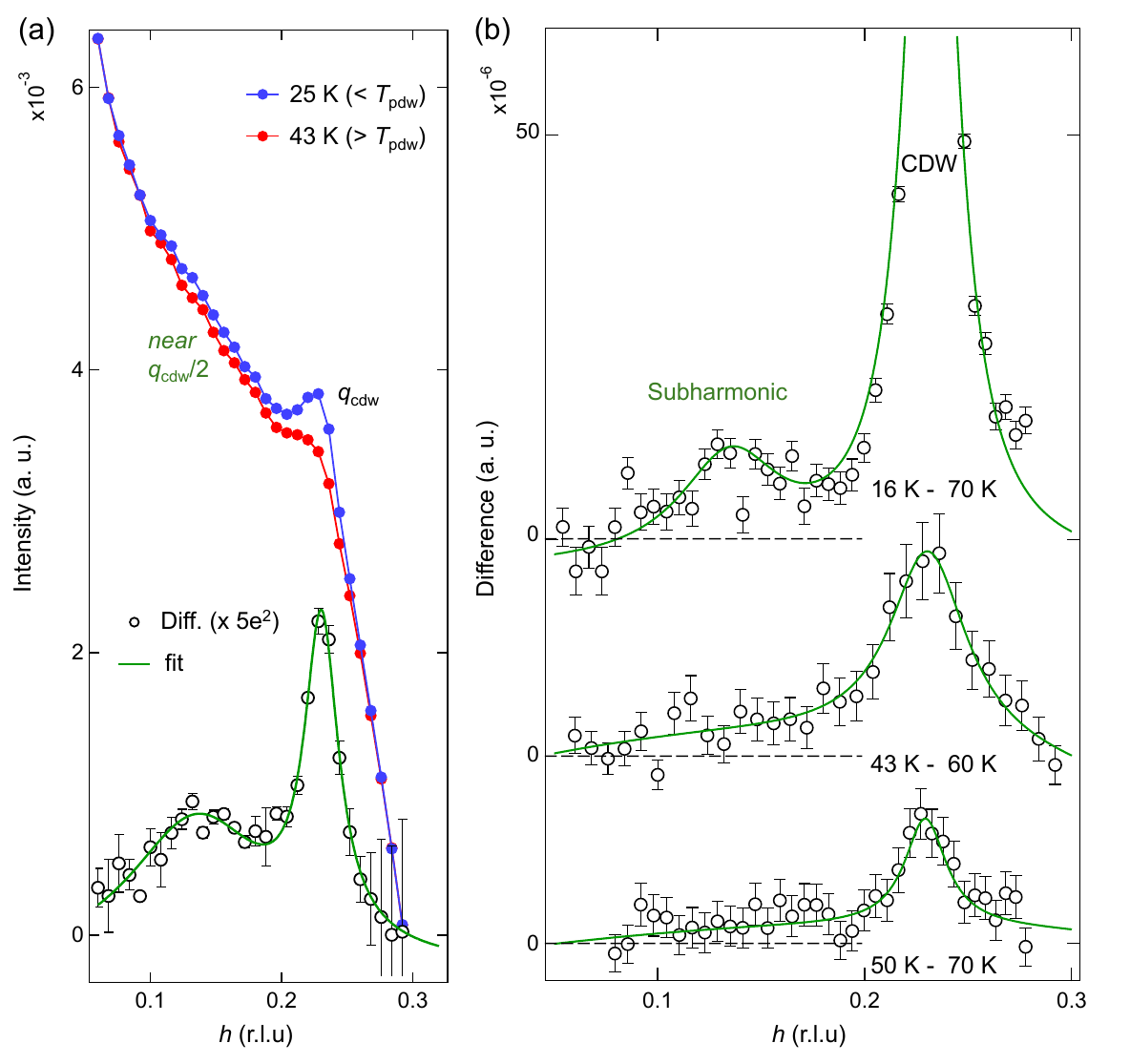}
\caption{Subharmonic scattering response in Fe-doped LSCO. (a) Representative momentum-dependent RSXS profiles measured along the CDW wave-vector direction below (25 K) and above (43 K) $T_{\rm pdw} \approx 32$~K. In addition to the dominant CDW peak near $q_{\rm cdw}$, an intensity feature is visible near $q_{\rm cdw}/2$, corresponding to the subharmonic wave vector. The difference profile (open symbols) between the two temperatures highlights the corresponding signal. For visibility, the difference profile in this panel is multiplied by 5$\times$10$^2$. This scaling is used only for display and is not used for quantitative extraction of the signal amplitude. (b) Temperature-difference RSXS profiles over selected temperature intervals across $T_{\rm pdw}$, illustrating the temperature dependence of the feature near $q_{\rm cdw}/2$ across $T_{\rm pdw}$. Solid lines represent simple fits used only as guides to the eye.}
\label{Fig4}
\end{center}
\end{figure}

Figure~\ref{Fig4}(a) shows representative momentum-dependent RSXS profiles measured below ($T = 25$~K) and above ($T = 43$~K) $T_{\rm pdw} \approx 32$~K. At 43~K, the scattering response is dominated by a weak, short-range CDW signal centered near $q_{\rm cdw}$ ($h \approx 0.24$~r.l.u.). Upon cooling below $T_{\rm pdw}$, the CDW scattering becomes more pronounced and sharper in momentum space, consistent with previous reports~\cite{Huang1}. However, even at comparable temperatures, the CDW contrast in LSCO remains relatively lower than that observed in Sr-doped LBCO [see Fig.~\ref{Fig2}(c)]. Despite this reduced CDW contrast, an excess intensity is still visible in the vicinity of $q_{\rm cdw}/2$, superimposed on the tail of the primary CDW scattering. The difference profile between the two temperatures highlights a corresponding signal at $h \approx 0.131 \pm 0.007$~r.l.u.\ [bottom plot in Fig.~\ref{Fig4}(a)]. Difference profiles measured at negative momentum transfer are consistent with those at positive momentum transfer within the statistical limits of the experiment ~\cite{Supple}.

%%%%%%%%%%%%%%%%Fig4%%%%%%%%%%%%%%%%%%%%%%%%%%%%%%
%%%%%%%%%%%%%%%%%%%%%%%%%%%%%%%%%%%%%%%%%%%%%%

Figure~\ref{Fig4}(b) further examines the temperature dependence of this feature by employing temperature-difference RSXS profiles constructed from several temperature intervals across $T_{\rm pdw}$. Given the relatively low CDW contrast in this system, such differential analysis provides a sensitive means to isolate subtle temperature-dependent changes that are not readily apparent in the raw profiles. The resulting difference profiles show that a discernible signal near $q_{\rm cdw}/2$ appears only when the temperature is below $T_{\rm pdw} \approx 32$~K. By contrast, temperature differences taken above $T_{\rm pdw}$ do not yield a comparable feature in this momentum region. This behavior closely parallels that observed in Sr-doped LBCO, indicating that, despite the reduced strength of stripe order in LSCO, the emergence of the subharmonic response is governed by the same temperature criterion.

\section{Discussion}

Before discussing the implications of the observed subharmonic response, we first clarify the scope and limitations of the present experiment. The reported signal is extremely weak and was measured in a limited set of samples and scattering geometries. Therefore, the present data do not by themselves exclude all possible background-related, structural, or experimental contributions to the observed temperature-dependent intensity. In this sense, the observation should not be viewed as a definitive, standalone proof of pair-density-wave order or as a complete characterization of a new order parameter. Rather, it should be regarded as a reproducible resonant scattering response observed near $q_{\rm cdw}/2$ within the experimentally accessible momentum window. With this limitation in mind, we discuss below the physical implications that follow if this response is intrinsic to the stripe-ordered, layer-decoupled regime of La-based cuprates.

Under this cautious interpretation, the present results have several implications for the phenomenology of intertwined superconducting and density-wave correlations in La-based cuprates. Taken together, the results presented here establish that a subharmonic scattering response near $q_{\rm cdw}/2$ can be observed in the temperature regime associated with interlayer decoupling in La-based cuprates. This behavior is found both in Sr-doped LBCO, where stripe order is well developed, and in Fe-doped LSCO, where stripe correlations are comparatively weaker. The consistency of this phenomenology across materials with markedly different stripe strengths indicates that the subharmonic response is tied to a common, temperature-defined physical regime, rather than arising solely from strong stripe order.

From a physical standpoint, this common regime can be discussed within the broader framework of intertwined orders in cuprates, in which the onset of interlayer decoupling marks a reorganization of charge, spin, and superconducting correlations. In stripe-ordered systems, charge and spin modulations provide a spatially modulated background that can coexist with superconducting correlations. These stripe correlations are generally understood to arise from spatially modulated electronic order within an otherwise homogeneous crystal, rather than from macroscopic phase separation. Within such a picture, the appearance of a scattering feature near $q_{\rm cdw}/2$ may be viewed as a momentum-space manifestation of this reorganization as the system enters the layer-decoupled regime. In this context, the observed scattering response near $q_{\rm cdw}/2$ may be discussed as reflecting underlying subharmonic correlations that emerge within this intertwined stripe--superconducting background.

At the same time, it is important to emphasize the scope of the present interpretation. The observed phenomenology does not require invoking a standalone ordered phase, but is naturally compatible with scenarios in which superconducting correlations are intertwined with preexisting stripe order. Although the scattering profiles exhibit peak-like features, the present RSXS measurements are inherently sensitive to momentum-resolved responses of buried electronic states and therefore do not aim to extract an order parameter from peak intensity or width, nor do they directly resolve the microscopic real-space structure of the underlying correlations. The present measurements do not provide a full $H$, $K$, and $L$ characterization of the subharmonic response. Accordingly, our conclusion is restricted to the temperature-dependent subharmonic scattering response observed within the experimentally accessible momentum window. Instead, the measurements document reproducible, momentum-resolved modifications of the scattering profile that evolve consistently within the same temperature-defined regime across different compounds, thereby constraining the phenomenology. Future complementary high-energy x-ray scattering over a wider momentum range could help clarify the momentum-space structure of this response. Further clarification of the microscopic origin will require complementary, feature-sensitive approaches capable of resolving symmetry, phase, and energy dependence.

%%%%%%%%%%%%%%%%%%%%%%%%%%%%%%%%%%%%%%%%%%%%%%

In summary, we used resonant soft x-ray scattering to identify a bulk-sensitive scattering signature of subharmonic charge-density-wave correlations in La-based cuprates. This response emerges reproducibly within a restricted thermodynamic regime associated with interlayer decoupling. Our emphasis is placed on the regime-specific appearance of this subharmonic component, rather than on peak-based order-parameter claims. Observing the same signature in two independent materials within the stripe-ordered, layer-decoupled regime provides an important experimental benchmark for future studies of intertwined superconducting states in high-$T_{\rm c}$ cuprates.

%%%%%%%%%%%%%%%%%%%%%%%%%%%%%%%%%%%%%%%%%%%%%%

\section*{Acknowledgments}

We thank John M. Tranquada, Peter Abbamonte, and David Hawthorn for insightful discussions and T. Masuda, S. Asai, and M. Ohkawara for supporting the neutron measurements at HER and TOPAN. All soft X-ray experiments were carried out at the SSRL (beamline 13-3), SLAC National Accelerator Laboratory, supported by the U.S. Department of Energy, Office of Science, Office of Basic Energy Sciences under Contract No. DE-AC02-76SF00515. The neutron scattering experiments were performed under the Joint-Use Research Program for Neutron Scattering, Institute for Solid State Physics, University of Tokyo, at the Japan Research Reactor JRR-3 (Program Nos. 21527 and 22402). We acknowledge the support from Center of Neutron Science for Advanced Materials, Institute for Materials Research, Tohoku University. S.A.K. acknowledges the support by Department of Energy, Office of Basic Energy Sciences, under contract no. DE-AC02-76SF00515 at Stanford. M.F. is supported by Grant-in-Aid for Scientific Research(S) (GrantNo.21H04987). %\\
%%%%%%%%%%%%%%%%%%%%%%%%%%%%%%%%%%%%%%%%%%%%%%



%

%\bibliography{prb_refs_fixed}

\end{document}